\documentclass[letter]{aa} % for the letters 
\usepackage{graphicx}
\usepackage{txfonts}
\usepackage{natbib}
\titlerunning{Angular momentum transport by the Tayler instability}

\begin{document} 

   \title{Rotation in stellar interiors: General formulation and an asteroseismic-calibrated transport by the Tayler instability}

\author{P. Eggenberger\inst{1}
\and F.D. Moyano\inst{1}
\and J.W. den Hartogh\inst{2}}

\institute{D\'epartement d'Astronomie, Universit\'e de Gen\`eve, Chemin Pegasi 51, CH-1290 Versoix, Switzerland \\     
\email{patrick.eggenberger@unige.ch}
\and
Konkoly Observatory, Research Centre for Astronomy and Earth Sciences, Konkoly Thege Mikl\'{o}s \'{u}t 15-17, H-1121 Budapest, Hungary
}
   \date{Received; accepted}

% \abstract{}{}{}{}{} 
% 5 {} token are mandatory
 
  \abstract
   {Asteroseismic measurements of the internal rotation of evolved stars indicate that at least one unknown efficient angular momentum (AM) transport mechanism is needed in stellar radiative zones in addition to hydrodynamic transport processes.}
   {We investigate the impact of AM transport by the magnetic Tayler instability as a possible candidate for such a missing physical mechanism.}
   {We derived general equations for AM transport by the Tayler instability to be able to test different versions of the Tayler-Spruit (TS) dynamo by comparing rotational properties of these models with asteroseismic constraints available for sub-giant and red giant stars.}
   {These general equations highlight, in a simple way, the key role played by the adopted damping timescale of the azimuthal magnetic field on the efficiency of the resulting AM transport. Using this framework, we first show that the original TS dynamo provides an insufficient coupling in low-mass red giants that have a radiative core during the main sequence (MS), as was found previously for more massive stars that develop a convective core during the MS. We find that the core rotation rates of red giant branch (RGB) stars predicted by models computed with various prescriptions for the TS dynamo are nearly insensitive to the adopted initial rotation velocity. We then derived a new calibrated version of the original TS dynamo and find that the damping timescale adopted for the azimuthal field in the original TS dynamo has to be increased by a factor of about 200 to correctly reproduce the core rotation rates of stars on the RGB. This calibrated version predicts no correlation of the core rotation rates with the stellar mass for RGB stars in good agreement with asteroseismic observations. Moreover, it correctly reproduces the core rotation rates of clump stars similarly to a revised prescription proposed recently. Interestingly, this new calibrated version of the TS dynamo is found to be in slightly better agreement with the core rotation rates of sub-giant stars, while simultaneously better accounting for the evolution of the core rotation rates along the RGB compared to the revised dynamo version. These results were obtained with both the Geneva and the MESA stellar evolution codes. 
}
{}

   \keywords{Stars: rotation -- Stars: magnetic field -- Stars: interiors -- Stars: oscillations}

   \maketitle
%
%-------------------------------------------------------------------

\section{Introduction}
\label{intro}

Asteroseismic observations of sub-giant and red giant stars have led to the determination of the internal rotation of these evolved stars \cite[][]{bec12, deh12, mos12, deh14, deh15, dim16, deh17, tri17, geh18, tay19, deh20}. These data are of prime interest for the modelling of angular momentum (AM) transport in stellar interiors. Comparisons with predictions of rotating stellar models solely accounting for hydrodynamic transport processes have shown that AM transport by meridional circulation and the shear instability provides an insufficient coupling to explain the internal rotation of red giants \citep{egg12_rg,cei13,mar13}. This indicates that at least one efficient AM transport process is missing in the radiative zone of these stars.

Fortunately, asteroseismic measurements of evolved stars can be used to precisely constrain the efficiency of the undetermined AM transport mechanism(s) during the post-main sequence (poMS) phase independently from any assumptions made for the modelling of AM transport or braking by magnetised winds on the main sequence (MS) \citep{egg17,egg19,moy22}. Magnetic transport mechanisms are prime candidates for explaining these asteroseismic measurements. Large-scale fossil magnetic fields could first be invoked to ensure uniform rotation in the radiative zones of evolved stars. Together with the assumption of radial differential rotation in their convective envelopes, this could explain the rotation rates observed in the core of red giants \citep{kis15, tak21}. However, detailed asteroseismic modellings of red giant branch (RGB) stars disfavour such a uniform rotation profile in the radiative interior of these stars \citep{kli17, fel21}.

A second possibility for efficient AM transport in stellar radiative zones is related to magnetic instabilities. The Tayler instability \citep{tay73} combined with the winding-up of a weak field by differential rotation is particularly interesting in this context \citep{spr02}. While this magnetic transport process  -- referred to as the Tayler-Spruit (TS) dynamo -- provides a physical explanation for the internal rotation of the Sun \citep{egg05_mag,egg19_sun}, it has been shown to be not efficient enough to reproduce the low core rotation rates of sub-giant and red giant stars \citep{can14,den19}. Recently, \citet{ful19} proposed a revised prescription for AM transport by the Tayler instability that predicts lower core rotation rates for sub-giant and red giant stars in better general agreement with observed values. Comparisons with asteroseismic data of evolved stars, however, have revealed that this revised prescription could not fully reproduce these observational constraints \citep{egg19_full,den20}. 

In this context, it is important to investigate, in more detail, the AM transport by the Tayler instability in the interior of evolved stars. We thus begin by deriving a general framework for AM transport by the Tayler instability that encompasses the original TS dynamo of \citet{spr02} as well as the revised prescription of \citet{ful19}. Based on these equations, we study the global features of these models and compare them to the asteroseismic constraints available for evolved stars. We then address the question of the asteroseismic calibration of the original TS dynamo. Indeed, previous comparisons between models computed with the original TS dynamo and observational constraints have always been carried out without any calibration parameters \citep[see e.g.][]{egg05_mag,can14,den19}, while such a parameter was introduced in the revised version of \citet{ful19}. 

The general equations for AM transport by the Tayler instability are described in Sect.~\ref{general_eqs} and the global rotational properties of these stellar models are compared to asteroseismic measurements in Sect.~\ref{results}. The calibrated version of the TS dynamo is discussed in Sect.~\ref{TScal}, while the conclusion is given in Sect.~\ref{conclusion}.

\section{General equations for AM transport by the Tayler instability}
\label{general_eqs}

We first discuss the basic equations for AM transport by the Tayler instability in stellar radiative zones. 
We begin by recalling that, in the context of AM transport by the TS dynamo, the azimuthal and radial components of the magnetic fields are given by \citep{spr02,ful19}
 \begin{equation}
 B_{\varphi}= (4 \pi \rho)^{\frac{1}{2}} \; r \; \omega_{\mathrm{A}} \quad \mathrm{and} \quad
 B_ {\mathrm{r}} \cong  B_{\varphi} \; \frac{\omega_{\mathrm{A}}}{N_{\rm eff}}  \; ,
 \label{champ}
 \end{equation}
\noindent with $r$ being the radius, $\rho$ the density, 
 $\omega_{\mathrm{A}}$ the corresponding Alfv\'en frequency, and $N_{\rm eff}$ the effective Brunt-V\"{a}is\"{a}l\"{a} frequency, which accounts for the reduction of the stabilising effect of the entropy gradient by thermal diffusion
 \begin{equation}
 N^2_{\rm eff}= \frac{\eta}{K} N_T^2 + N_{\mu}^2 \; ,
 \end{equation}
 
\noindent where $N_T$ and $N_{\mu}$ denote the thermal and chemical composition components of the Brunt-V\"{a}is\"{a}l\"{a} frequency, while $K$ and $\eta$ are the thermal and magnetic diffusivities.

The characteristic timescale $ \tau_{\mathrm{amp}}$ on which the radial component of the magnetic field is amplified into an azimuthal component of the same amplitude as the one of the existing field is given by
 \begin{equation}
  \tau_{\mathrm{amp}} r   \left| \frac{\partial \Omega}{\partial r} \right| =  \frac{B_{\varphi}}{B_{\mathrm{r}}} \; ,
  \label{tau_amp_v1}
\end{equation}

\noindent with $\Omega(r)$ being the angular velocity. Introducing the shear parameter $q= \left| \frac{\partial \ln \Omega}{\partial \ln r} \right| $ and using Eq.~\ref{champ}, the amplification timescale becomes
 \begin{equation}
  \tau_{\mathrm{amp}}=  \frac{N_{\rm eff}}{\omega_{\mathrm{A}} \Omega q} \; .
  \label{tau_amp}
\end{equation}

The main difference between the original prescription for transport by the Tayler instability proposed by \cite{spr02} and the revised prescription proposed by \cite{ful19} is related to the saturation of the instability. In the original case, the damping timescale of the azimuthal magnetic field follows directly from the physics of the Tayler instability; it is then considered as the inverse of the growth rate of the instability. The typical growth rate of the Tayler instability is $\omega_{\mathrm{A}}$ for slow rotation, that is to say for $\Omega << \omega_{\mathrm{A}} $. However, this condition is not easily met in stellar interiors (even in the case of a slowly rotating star such as the Sun) so that the case of fast rotation, that is $\Omega >> \omega_{\mathrm{A}} $, generally has to be considered. This leads to the reduction of the growth rate of the instability by the factor $\omega_{\mathrm{A}} / \Omega$ as a result of the Coriolis force \citep{pit85,spr99}. In a revised prescription, \cite{ful19} instead introduced a turbulence formalism that focusses on energy dissipation. They then find a lower energy dissipation rate than in the original prescription, which then results in higher amplitudes of the magnetic fields and hence a more efficient AM transport. This difference with the formulation by \cite{spr02} can be expressed by a damping timescale of the azimuthal magnetic field that is increased by a factor $(\Omega / \omega_{\mathrm{A}})^2$ compared to the original prescription. Consequently, this general damping timescale can be written in a compact form as

 \begin{equation}
  \tau_{\mathrm{damp}}= C_{\rm T} \; \frac{1}{\omega_{\mathrm{A}}} \left( \frac{\Omega}{\omega_{\mathrm{A}}} \right)^n  \; ,
  \label{tau_damp}
\end{equation}
\noindent with $n=1$ for the original TS dynamo and $n=3$ for the revised prescription. We introduce here a dimensionless calibration parameter $C_{\rm T}$ to account for uncertainties on this adopted timescale.
By requiring that both timescales be equal, one then obtains the following:

\begin{equation}
\left( \frac{\omega_{\mathrm{A}}}{\Omega} \right)^n = C_{\rm T} \; q \; \frac{\Omega}{N_{\rm eff}}  \; .
\label{omega_A_n}
\end{equation}
\noindent

To express the corresponding AM transport, we write the effective viscosity $\nu_{\rm T}$ for the vertical transport by the Tayler instability \citep{spr02}:

 \begin{equation}
 \nu_{\rm T} = \frac{B_{\mathrm{r}} B_{\varphi}}{4 \pi \rho q \Omega} =  \frac{1}{4 \pi \rho q \Omega} \left(\frac{\omega_{\mathrm{A}}}  {N_{\rm eff}}\right) B_{\varphi}^2 =
  \; \frac{\Omega \; r^2}{q} \;
 \left( \frac{\omega_{\mathrm{A}}}{\Omega}\right)^3 \; 
\left(\frac{\Omega}{N_{\rm eff}}\right) \; .
 \label{nu}
 \end{equation}

\noindent
Using Eq.~\ref{omega_A_n}, we then obtain the general expression for the viscosity associated with\ AM transport by the Tayler instability:

 \begin{equation}
 \nu_{\rm T} =
  \; \frac{\Omega \; r^2}{q} \;
 \left(  C_{\rm T} \; q \; \frac{\Omega}{N_{\rm eff}}  \right)^{3/n} \; 
\left(\frac{\Omega}{N_{\rm eff}}\right) \; .
 \label{nu_generale}
 \end{equation}

When $n=1$, the general Eq.~\ref{nu_generale} leads to the viscosity corresponding to the original prescription for the TS dynamo (Eq.~31 of \cite{spr02}):

 \begin{equation}
 \nu_{\rm TS}= C_{\rm T}^3 r^2 \Omega q^2 \left(\frac{\Omega}{N_{\rm eff}}\right)^4  \; ,
 \label{nu_TS_Nmu}
 \end{equation}

\noindent
while $n=3$ corresponds to the viscosity associated with the prescription proposed by \cite{ful19}
(Eq.~35 of \cite{ful19}):
\begin{equation}
 \nu_{\rm F}= C_{\rm T} r^2 \Omega \left(\frac{\Omega}{N_{\rm eff}}\right)^2 \; . 
 \label{nu_F}
 \end{equation}
\noindent
Comparing this equation with Eq.~35 from \cite{ful19}, one notes that the calibration constant $C_{\rm T}$ introduced here for the damping timescale corresponds to $\alpha^3$, with $\alpha$ being the dimensionless calibration parameter used in \cite{ful19} to account for uncertainties in the saturated Alfv\'en frequency.

The general expression for the minimum value of the shear parameter $q_{\rm min}$ required for the magnetic process to operate is obtained by introducing the critical value for the azimuthal field to become unstable to the Tayler instability 
\begin{equation}
\left(\frac{\omega_{\mathrm{A}}}{\Omega}\right)^4  =  \frac{N_{\rm eff}^2}{\Omega^2} \;
\frac{\eta}{r^2 \; \Omega}   \; ,
\label{premier}
\end{equation}
\noindent
in Eq. \ref{omega_A_n}. We then obtain the following general equation for the minimum value of the shear parameter:

\begin{equation}
q_{\rm min, T} = C_{\rm T}^{-1} \left(\frac{N_{\rm eff}}{\Omega}\right)^{(n+2)/2} \left(\frac{\eta}{r^2 \Omega}\right)^{n/4} \; .
\label{qmin_generale}
\end{equation}
As for the expression of the viscosity associated with the transport by the Tayler instability, the case $n=1$ corresponds to the original TS dynamo (Eq.~26 of \cite{spr02}):

\begin{equation}
q_{\rm min, TS } = C_{\rm T}^{-1} \left(\frac{N_{\rm eff}}{\Omega}\right)^{3/2} \left(\frac{\eta}{r^2 \Omega}\right)^{1/4} = C_{\rm T}^{-1} \left(\frac{N_{\rm eff}}{\Omega}\right)^{7/4} \left(\frac{\eta}{r^2 N_{\rm eff}}\right)^{1/4}\; .
\label{qmin_TS}
\end{equation}
Similarly, $n=3$ in Eq.~\ref{qmin_generale} leads to the expression derived by \cite{ful19} (Eq. 36):
\begin{equation}
q_{\rm min, F } = C_{\rm T}^{-1} \left(\frac{N_{\rm eff}}{\Omega}\right)^{5/2} \left(\frac{\eta}{r^2 \Omega}\right)^{3/4} \; .
\label{qmin_F}
\end{equation}

\section{Global properties of rotating models with AM transport by the Tayler instability}
\label{results}

Rotating models are computed with the Geneva stellar evolution code \citep{egg08} using the general theoretical framework for transport by the Tayler instability described above. AM transport by the Tayler instability is then taken into account through the viscosity $\nu_{\rm T}$ given in Eq.~\ref{nu_generale} when the shear parameter $q$ is larger than the threshold $q_{\rm min, T}$ given in Eq.~\ref{qmin_generale}. In addition to magnetic AM transport, transport by meridional circulation and the shear instability is taken into account. With the assumption of shellular rotation \citep{zah92}, the following equation is then solved for AM transport in radiative zones simultaneously to the evolution of the star:
\begin{equation}
  \rho \frac{{\rm d}}{{\rm d}t} \left( r^{2}\Omega \right)_{M_r} 
  =  \frac{1}{5r^{2}}\frac{\partial }{\partial r} \left(\rho r^{4}\Omega
  U(r)\right)
  + \frac{1}{r^{2}}\frac{\partial }{\partial r}\left(\rho (D_{\rm shear}+\nu_{\rm T}) r^{4}
  \frac{\partial \Omega}{\partial r} \right) \, .
\label{transmom}
\end{equation}
\noindent We note that $U(r)$ is the radial dependence of the meridional circulation velocity in the radial direction \citep{mae98}. Transport of AM by the shear instability is accounted for with the coefficient $D_{\rm shear}$ of \citet{tal97}.

\begin{figure}[htb!]
\resizebox{\hsize}{!}{\includegraphics{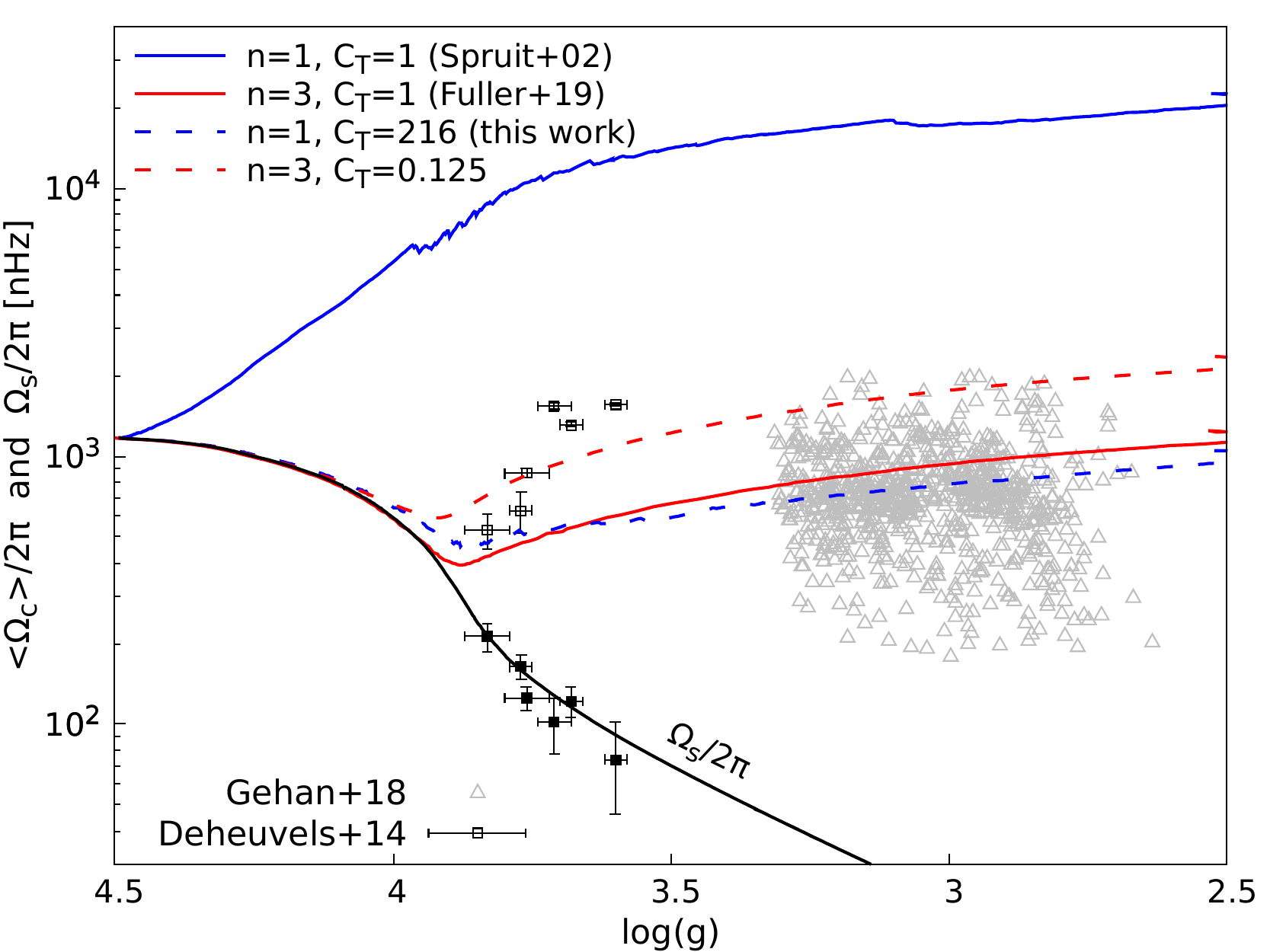}}
 \caption{Core rotation rates as a function of surface gravity for 1.1\,$M_{\odot}$ models computed with the general equations (Eqs.~\ref{nu_generale} and ~\ref{qmin_generale}) for AM transport by the Tayler instability with different values of $n$ and the calibration constant $C_{\rm T}$. The blue and red lines correspond to $n=1$ and $n=3$, respectively. Solid blue and red lines indicate models with $C_{\rm T}=1$. Dashed blue and red lines correspond to $C_{\rm T}=216$ and $C_{\rm T}=0.125$, respectively. The solid black line indicates the evolution of the surface rotation rate.}
  \label{omc_1p1}
\end{figure}

To study the global rotational properties of models of evolved stars accounting for AM transport by the Tayler instability, 1.1\,$M_{\odot}$ models are first computed for different values of $n$ and of the calibration constant $C_{\rm T}$. A solar chemical composition is assumed for these models, together with a solar-calibrated mixing-length parameter for convection. The initial velocity is chosen to obtain a surface rotation period of $\sim 60$ days at $\log g \sim 3.8$ that is needed to correctly reproduce the surface rotation rates of sub-giant stars observed by \citet{deh14} (see the black line in Fig.~\ref{omc_1p1}). 

The evolution of the core rotation rate of these models is shown in Fig.~\ref{omc_1p1}. For the same value of the calibration parameter $C_{\rm T}=1$ (solid lines), one clearly sees that increasing $n$ from one to three results in a more efficient AM transport that leads to a better global agreement with the core rotation rates observed for evolved stars. Recalling that $n=1$, $C_{\rm T}=1$ corresponds to the original TS dynamo, we thus confirm previous findings that this mechanism does not predict a sufficient coupling to correctly account for the low core rotation rates of red giants \citep{can14,den19}. While this result was only obtained previously for stars massive enough to have a convective core during the MS, we show here that the same conclusion is reached for lower mass stars with a radiative core during the MS. 

Figure~\ref{omc_1p1} also shows that the more efficient coupling associated with $n=3$ leads to a much better agreement with the core rotation rates of red giants for $C_{\rm T}=1$, as was found by \cite{ful19}. However, too low core rotation rates are then predicted for sub-giants compared to asteroseismic constraints \citep{egg19_full}. A change in the calibration parameter is then needed to better account for the core rotation rates of sub-giant stars as illustrated by the $C_{\rm T}=0.125$ case (recalling that $C_{\rm T}=\alpha^3$, this corresponds to $\alpha=0.5$, dashed red line in Fig.~\ref{omc_1p1}). In a similar way, an increase in the calibration parameter for the $n=1$ case enables one to correctly account for the core rotation rates of red giants as shown by the $C_{\rm T}=216$ case (dashed blue line in Fig.~\ref{omc_1p1}). Interestingly, this case is also in slightly better agreement with the core rotation rates of sub-giant stars, while  simultaneously being in better agreement with the core rotation rate of red giants compared to the $n=3$ and $C_{\rm T}=1$ case.

\begin{figure}[htb!]
\resizebox{\hsize}{!}{\includegraphics{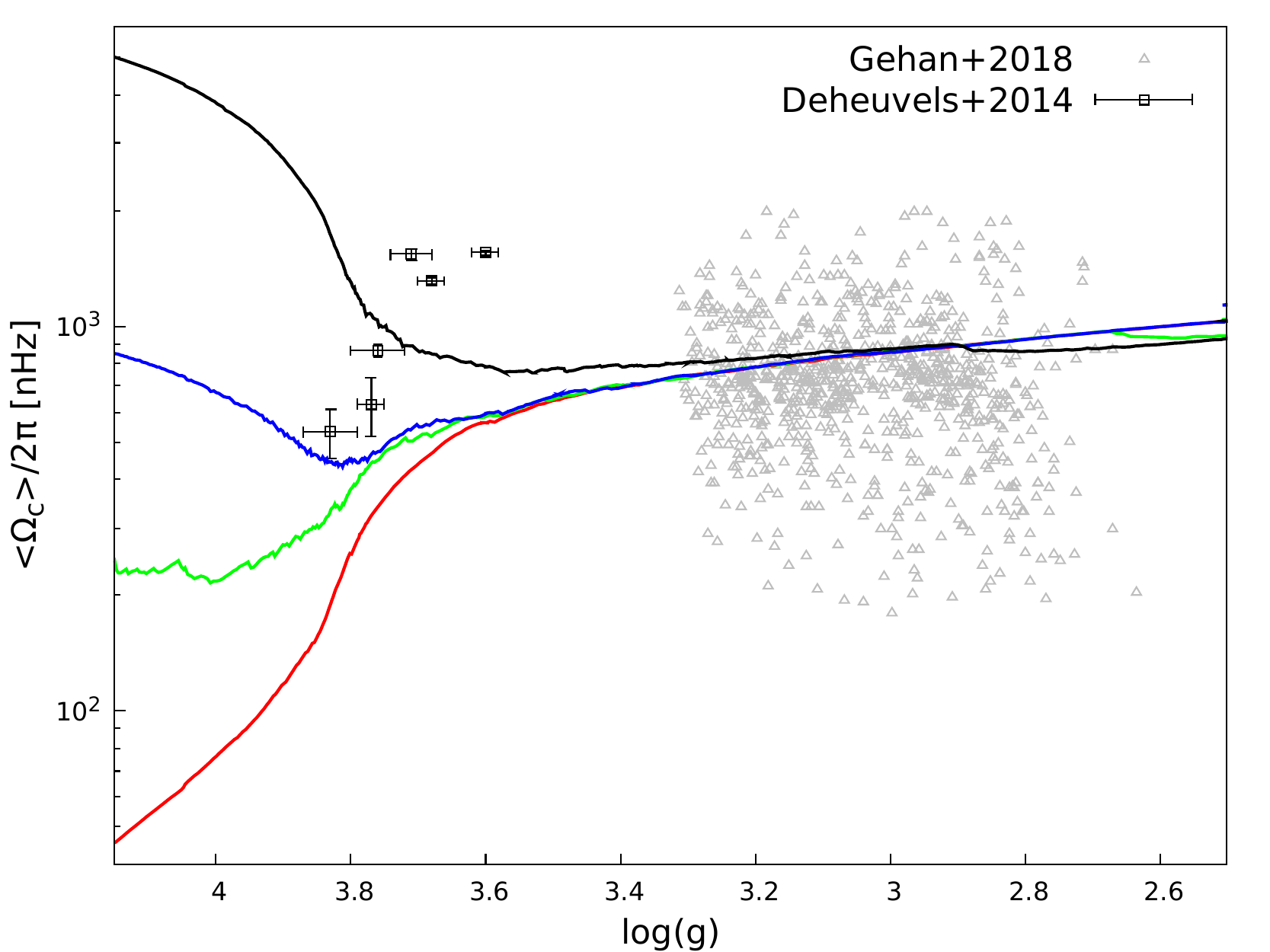}}
 \caption{Core rotation rates as a function of surface gravity for 1.2\,$M_{\odot}$ models with $n=1$ and $C_{\rm T}=216$ computed for different initial rotation velocities. Red, green, blue, and black lines indicate the core rotation rates for models with a surface rotation period of about 3300, 300, 60, and 12 days at $\log g \sim3.8$, respectively.}
  \label{omc_1p2}
\end{figure}

This $n=1$, $C_{\rm T}=216$ case corresponds to a new asteroseismic-calibrated version of the original TS dynamo. The parameter $C_{\rm T}$ is introduced here to account for uncertainties on the adopted timescale for the damping of the azimuthal field (while the $\alpha$ parameter in \cite{ful19} was introduced for the saturated value of $\omega_{\mathrm{A}}$, hence the relation $C_{\rm T}=\alpha^3$). We thus find that the damping timescale adopted for the azimuthal field in the original TS dynamo has to be increased by a factor of about 200 to correctly reproduce the asteroseismic data of evolved stars. Ideally, one would expect a value of $C_{\rm T}$ closer to unity for a well-defined physical process. We however recall here that this timescale is known to be quite approximated as mentioned by \citet{spr02} who indicated that the original estimates of the dynamo process were made by neglecting all multiplying factors of order unity and that these factors could sometimes compound to rather large numbers. A first uncertainty is related to the exact value of the growth rate of the Tayler instability which seems to be somewhat smaller than the adopted value of $\omega_{\mathrm{A}}$ \citep[e.g.][]{gol19}. In the same way, the correction to this growth rate in the case of fast rotation (i.e. $\Omega >> \omega_{\mathrm{A}}$) introduces another uncertainty that also seems to overestimate its value \citep{iba15}. Another source of uncertainty is related to the fact that these estimates of the growth rate are only based on the fastest growing non-axisymmetric $m=1$ Tayler mode, while the small-scale dynamo process could perfectly be dominated by modes with different values of $m$ \citep[see e.g.][]{iba15}. In this context, the damping timescale adopted in the original TS dynamo seems to correspond more to a minimal than an exact value; it is thus interesting to be able to constrain its value from asteroseismic data and it is then not surprising to deduce a longer timescale in this way. Of course, it is difficult to speculate whether such a large increase of about two orders of magnitude is really physically motivated in the framework of the original TS dynamo and we can only await numerical simulations performed under more realistic stellar conditions to obtain some answers on this point. 

Another possibility is to change the framework of the original TS dynamo somewhat to be able to reproduce such an increase of about 200 in the damping timescale while keeping a growth rate equal to $\omega_{\mathrm{A}}^2/\Omega$. This corresponds to the revision of \cite{ful19} which, as shown above with Eq.~\ref{tau_damp}, leads to an increase by a factor of $(\Omega/\omega_{\mathrm{A}})^2$ in the damping timescale of the azimuthal field compared to the original prescription. With $\Omega$ being about one order of magnitude larger than $\omega_{\mathrm{A}}$ in the hydrogen-burning shell of a red giant \citep[see e.g. Fig.~2 of][]{ful19}, this factor then results in the required change of about two orders of magnitude for this timescale, thereby providing an interesting physical explanation for this value.

\section{An asteroseismic-calibrated version of the Tayler-Spruit dynamo}
\label{TScal}

A key feature of the revised version of the dynamo with $n=3$ found by \citet{ful19} is that the core rate rotation during the RGB phase is almost insensitive to the initial rotation velocity of the models \citep[see Fig.~4 of][]{ful19}. To investigate whether this result corresponds to a specific property of the revised formulation of \citet{ful19} or to a general feature of models accounting for AM transport by the Tayler instability, we computed models for the $n=1$, $C_{\rm T}=216$ case with different initial rotation rates. This is illustrated in Fig.~\ref{omc_1p2} for models with a mass of 1.2\,$M_{\odot}$ and a surface rotation period of about 12, 60, 300, and 3300 days during the sub-giant phase (i.e. at $\log g \sim 3.8$) for the black, blue, green, and red lines, respectively. While differences in the core rotational properties are important on the MS and rapidly decrease during the sub-giant phase, similar core rotation rates are obtained on the RGB regardless of the initial rotation velocity adopted. The fact that core rotation rates on the RGB are nearly insensitive to initial rotation velocities is then found to be a global feature of AM transport by the Tayler instability as described by Eqs.~\ref{nu_generale} and \ref{qmin_generale}. 

\begin{figure}[htb!]
\resizebox{\hsize}{!}{\includegraphics{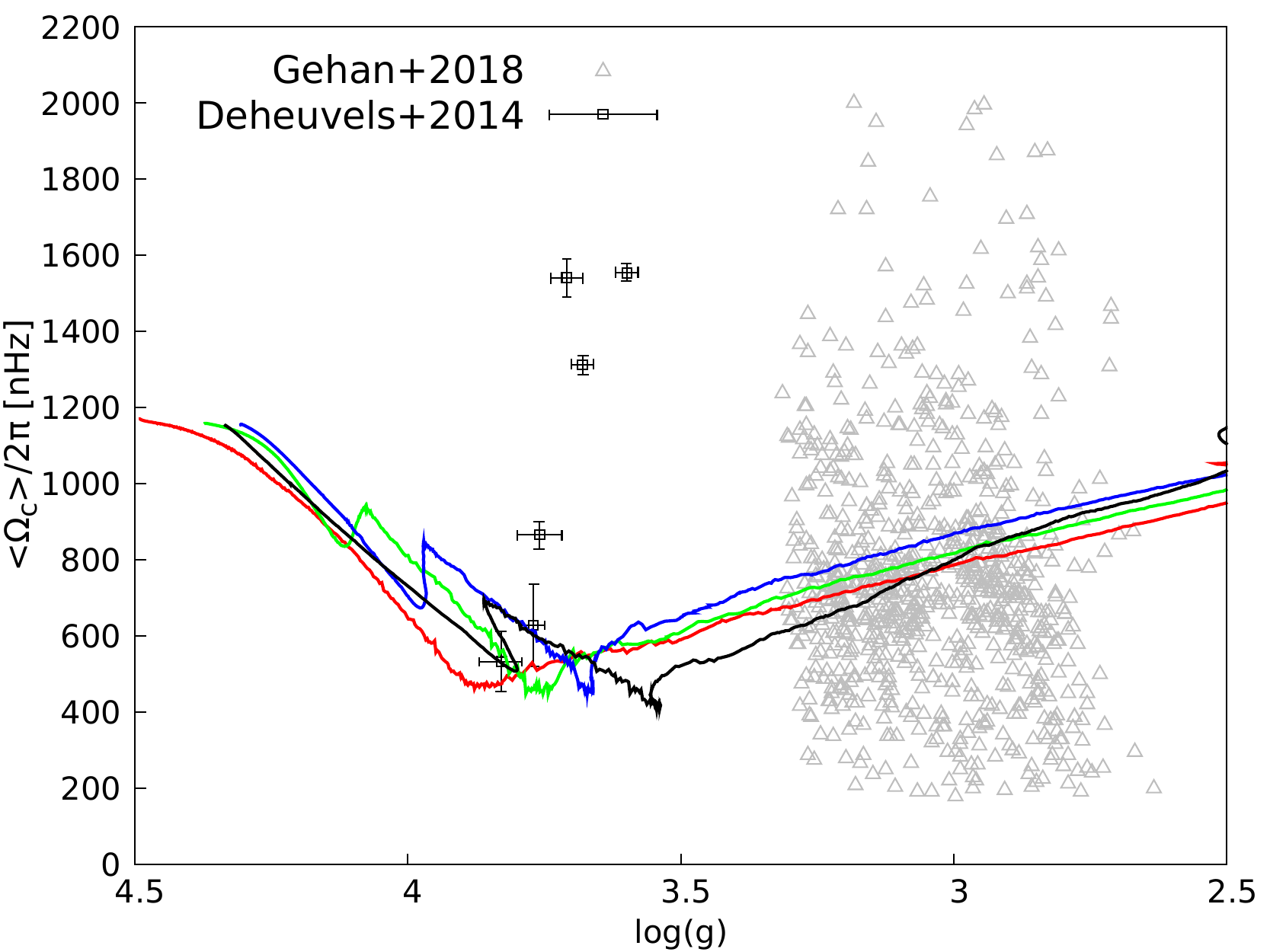}}
 \caption{Same as Fig.~\ref{omc_1p2}, but for models with the same initial rotation period yet different masses. Red, green, blue, and black lines indicate a mass of 1.1, 1.3, 1.5, and 1.7\,$M_{\odot}$, respectively.}
  \label{omc_mass}
\end{figure}

Another important constraint deduced from asteroseismic measurements of a large number of red giants is that there is no correlation between the core rotation rates and the stellar mass \citep{mos12,geh18}. As discussed in the previous section, the $n=1$, $C_{\rm T}=216$ case is promising to reproduce the observations for red giants with masses of 1.1 and 1.2\,$M_{\odot}$. We now investigate the trend with the stellar mass by computing models with $n=1$, $C_{\rm T}=216$ that share the same initial rotation period, but have masses of 1.3, 1.5, and 1.7\,$M_{\odot}$. Figure~\ref{omc_mass} shows that models with different masses exhibit similar core rotation rates on the RGB, which correctly reproduce the asteroseismic constraints. We thus find that the calibrated version of the original TS dynamo is able to account for the absence of correlation between core rotation rates and stellar masses.   

\begin{figure}[htb!]
\resizebox{\hsize}{!}{\includegraphics{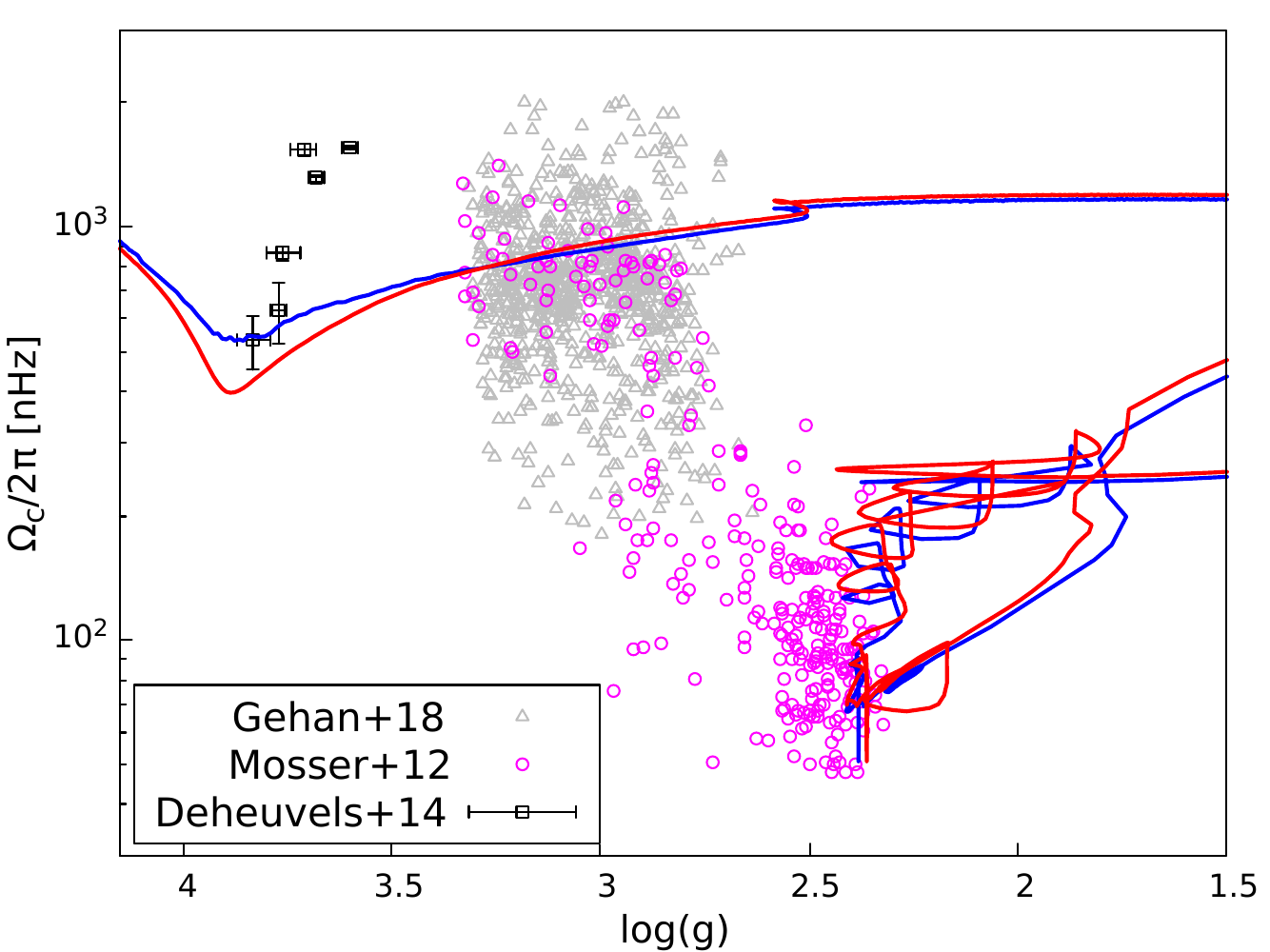}}
 \caption{Core rotation rates as a function of the surface gravity for 1.1\,$M_{\odot}$ models of sub-giant and red giant stars computed with the MESA code. The red and blue lines correspond to the revised version of the dynamo by \citet[][]{ful19} (i.e. $n=3$ and $C_{\rm T}=1$) and the asteroseismic-calibrated version of the TS dynamo ($n=1$ and $C_{\rm T}=216$), respectively.}
  \label{omc_clump}
\end{figure}

All results discussed previously have been obtained from models computed with the Geneva stellar evolution code. Comparisons between the MESA \citep[][]{pax11,pax13,pax15,pax18,pax19} and the Geneva codes have already shown that similar rotational properties are obtained for the revised version of the TS dynamo, that is the $n=3$ case \citep{egg19_full}. We checked that this is also the case for the asteroseismic-calibrated version of the TS dynamo by computing a MESA 1.1\,$M_{\odot}$ model with $n=1$, $C_{\rm T}=216$. This model is shown by the blue line in Fig.~\ref{omc_clump}, while the red line indicates the corresponding MESA model with $n=3$, $C_{\rm T}=1$. Comparing Figs.~\ref{omc_1p1} and \ref{omc_clump} confirms that similar core rotation rates were obtained with the Geneva and the MESA codes for the $n=1$, $C_{\rm T}=216$ model. The better agreement with the asteroseismic determinations of core rotation rates for sub-giant and red giant stars of the $n=1$, $C_{\rm T}=216$ case compared to the $n=3$, $C_{\rm T}=1$ case was also obtained with MESA. Moreover, Fig.~\ref{omc_clump} shows that, once the core rotation rates are correctly reproduced on the RGB, these magnetic models naturally predict core rotation rates for core-helium burning stars (magenta circles with $\log g$ lower than about 2.5 in Fig.~\ref{omc_clump}) in good agreement with the asteroseismic measurements of \citet{mos12}. This result was already obtained by \citet{ful19} for the revised version of the dynamo ($n=3$, $C_{\rm T}=1$) and we find that this is also the case for the calibrated version of the original TS prescription.

\section{Conclusion}
\label{conclusion}

General equations for AM transport by the Tayler instability that encompass the original TS dynamo \citep{spr02} and the revised one \citep{ful19} were derived 
first. Based on these equations, we then study -- within the same numerical scheme -- the global features of models accounting for AM transport by the Tayler instability and compare them to the asteroseismic constraints available for evolved stars.

We first show that the AM transport predicted by the original TS dynamo is not efficient enough in low-mass stars with a radiative core during the MS to correctly account for asteroseismic constraints. This finding complements previous studies that reached the same conclusion, but only for stars massive enough to have a convective core on the MS \citep{can14,den19}. We then address the question of the calibration of the TS dynamo and find that the damping timescale adopted for the azimuthal field in the original TS dynamo has to be increased by a factor of about 200 to correctly reproduce the core rotation rates of red giant stars.
We also show that this calibrated TS dynamo correctly reproduces the core rotation rates of clump stars similarly to the revised prescription of \citet{ful19}. Interestingly, this new calibrated version of the TS dynamo is found to be in slightly better agreement with the core rotation rates of sub-giant stars, while simultaneously better reproducing the evolution of the core rotation rates along the RGB compared to the revised dynamo version proposed by \citet{ful19}. We also show that the fact that the core rotation rate of a red giant is nearly insensitive to its initial rotation velocity is a general feature of models accounting for AM transport by the Tayler instability. Finally, we find that the new calibrated version of the TS dynamo predicts no correlation between the core rotation rates and the masses of RGB stars, which is in good agreement with asteroseismic observations.

\begin{acknowledgements}
PE and FDM have received funding from the European Research Council (ERC) under the European Union's Horizon 2020 research and innovation programme (grant agreement No 833925, project STAREX). JWdH has received funding from the ERC Consolidator Grant (Hungary) programme (RADIOSTAR, G.A. n. 724560).
\end{acknowledgements}

%-------------------------------------------------------------------

\bibliographystyle{aa} % style aa.bst
\bibliography{biblio} % your references Yourfile.bib

\end{document}